\documentclass[prb,twocolumn,showpacs,english]{revtex4}

\usepackage{amsmath}
\usepackage{graphicx}
\usepackage{amssymb}
\usepackage{times,mathptm,dsfont,amsbsy,fancyhdr}

\usepackage{babel}

\begin{document}

\title{Semiclassical theory of ballistic transport through chaotic
cavities with spin-orbit interaction}

\author{Jens Bolte}\email{jens.bolte@uni-ulm.de} 
\affiliation{Institut f\"ur Theoretische Physik, Universit\"at Ulm, 
Albert-Einstein-Allee 11, D-89069 Ulm, Germany}
\author{Daniel Waltner}\email{Daniel.Waltner@physik.uni-regensburg.de} 
\affiliation{Institut f\"ur Theoretische Physik, Universit\"at Regensburg, 
D-93040 Regensburg, Germany}

\date{\today}

\begin{abstract}
We investigate the influence of spin-orbit interaction on ballistic
transport through chaotic cavities by using semiclassical methods. 
Our approach is based on the Landauer formalism and the Fisher-Lee relations,
appropriately generalized to spin-orbit interaction, and a semiclassical
representation of Green functions. We calculate conductance coefficients 
by exploiting ergodicity and mixing of suitably combined classical
spin-orbit dynamics, and making use of the Sieber-Richter method and
its most recent extensions. That way we obtain weak anti-localization 
and confirm previous results obtained in the symplectic ensemble of Random 
Matrix Theory.
\end{abstract}

\pacs{73.23.-b,71.70.Ej,72.15.Rn,03.65.Sq,05.45.Mt}

\maketitle

\section{Introduction}
Ballistic transport through chaotic cavities realized as quantum dots 
in semiconductor heterostructures has been a central issue in mesoscopic 
physics for many years. The universal transport properties observed in
this context can be described on a phenomenological level by random
matrix theory~\cite{Bee97} (RMT). The same applies to disordered systems, 
where averages over impurities can be shown to be equivalent to random 
matrix averages. This not being possible for individual, clean cavities,
theoretical explanations of the RMT-connection have been provided
making use of semiclassical methods, which are based on the Landauer 
formalism~\cite{Lan57} and semiclassical representations of Green functions. 
This approach~\cite{Bar93} leads to questions that are closely analogous 
to problems arising in semiclassical explanations of universal spectral 
correlations in classically chaotic quantum systems. Recent progress in the 
latter context is based on the seminal work of Sieber and Richter~\cite{Ric01}
and its extensions~\cite{Mue04,Heu05,He07}. This method has been 
adapted~\cite{Sie02,Heus06,Heu07} to be able to successfully explain 
conductance coefficients, including the effect of weak localization, i.e., 
a decrease of conductance at zero magnetic field. Further studies have been 
devoted to analyses of the universality of conductance 
fluctuations~\cite{Bro05,Heu07,Ber07}, and of shot 
noise~\cite{Heu06,Sch03,Heu07}. (For an overview see, e.g., 
Ref.~\onlinecite{Heu07}). 

In the work mentioned transport properties were considered for ballistic,
non-relativistic electrons, neglecting their spin. In the emerging field 
of semiconductor based spin electronics~\cite{Fab04} (spintronics), however, 
one requires an efficient control of the spin dynamics associated with 
electrons in non-magnetic semiconductors. This purpose calls for an inclusion 
of spin-orbit interactions into studies of transport properties. In contrast 
to previous theories neglecting the spin, here one would expect appropriate
classical spin-orbit dynamics to produce weak anti-localization, 
i.e., an enhancement of the conductance at zero magnetic field. This 
prediction is also obtained on the phenomenological level provided by 
RMT, where a half-integer spin requires the symplectic, as opposed to the 
orthogonal, circular ensemble. On this ground one expects universal 
conductance fluctuations and shot noise also to be affected by the
presence of spin-orbit interactions~\cite{Bee97,Sav06}. A first semiclassical 
approach~\cite{Zai05} to these questions employs the semiclassical
representation of the Green function in spin-orbit coupling systems
derived in Ref.~\onlinecite{Bol99} and considers the first order of the 
semiclassical Sieber-Richter expansion. It, moreover, assumes a 
randomization of spin states, which is shown to be responsible for weak 
anti-localization.

In this paper our goal is to extend the results of Ref.~\onlinecite{Zai05}
to all orders of the Sieber-Richter expansion, and to base the semiclassical
estimates entirely on dynamical properties of suitably combined classical 
spin-orbit dynamics~\cite{Bol99a}. These then replace the randomization 
hypothesis of spin states made in the analytic part of 
Ref.~\onlinecite{Zai05}. 
In order to determine the spin contribution to transmission amplitudes we 
closely follow an analogous calculation introduced in the context of 
semiclassical explanations of spectral correlations in quantum graphs with 
spin-orbit couplings~\cite{Bol03,Bol05}. We also comment on shot noise and 
on the variance of conductance fluctuations.

As our model we consider a two dimensional cavity with two straight,
semi-infinite leads with hard walls. Apart from boundary reflections,
particles with mass $m$, charge $e$, and spin $s$ move freely within the 
leads and are subjected to a magnetic field and to spin-orbit interactions 
inside the cavity. Although the relevant case of electrons enforces the
spin to be $s=1/2$, we deliberately allow for general spin $s$. Below
this will allow us to point out characteristic differences between integer
and half-integer spin. The Hamiltonian governing the dynamics in the 
cavity reads
\begin{equation}
\label{eq:hamilton}
 \hat H = \frac{1}{2m} \left( \hat{\mathbf{p}}-\frac{e}{c}\mathbf{A}
 \left(\hat{\mathbf{x}}\right) \right)^{2} + \hat{\mathbf{s}}\cdot
 \mathbf{C}\left(\hat{\mathbf{x}},\hat{\mathbf{p}}\right)\ .
\end{equation}
Here $\mathbf{A}$ is the vector potential for an external magnetic field
and $\mathbf{C}$ contains all couplings of the translational degrees of 
freedom to the spin operator $\hat{\mathbf{s}}$. This may include Zeeman-, 
spin-orbit, Rashba-, or Dresselhaus-type couplings. Moreover, in order to 
model the hard walls we require Dirichlet conditions at the boundaries of 
the cavity and of the leads. 

The paper is organized as follows: Section~II is devoted to a generalization 
of the Landauer formalism and the Fisher-Lee relations to systems with 
spin-orbit interaction. Then we present semiclassical representations of
S-matrix elements in that case. In Section~III we first introduce 
ergodicity and mixing conditions that include a classical spin-orbit 
interaction. This is followed by our calculation of the conductance
in two ways: in the configuration-space and in the phase-space approach. 
In Sections~IV and V we then outline how our approach can be extended 
to calculate shot noise and conductance fluctuations, respectively. An 
Appendix contains a calculation whose result is central to the phase-space
approach employed in Section~III.
\section{Preliminaries}
We follow the usual approach to obtain semiclassical approximations
to transmission by employing the Landauer formalism \cite{Lan57} and 
introducing semiclassical representations for Green functions. In the 
absence of spin-orbit interactions this procedure is well 
established~\cite{Fis81,Bar89,Noe93}. Here we briefly describe the 
extensions required by the presence of spin-orbit interactions (see also 
Ref.~\onlinecite{Zai05}).
\subsection{Landauer formalism with spin}
The Landauer formalism provides a link between conductance coefficients,
as defined through
\begin{equation}
\label{eq:conductdef}
 I_{n}=\sum_{m}g_{nm}V_{m} \ ,
\end{equation}
and $S$-matrix elements. In (\ref{eq:conductdef}) the indices
label the leads, $V_{m}$ is the voltage applied at lead $m$ and $I_{n}$
is the current through lead $n$. Here the number of leads may be
arbitrary. An $S$-matrix element $S^{nm}_{\alpha_n\alpha'_m}$ is defined 
as the transition amplitude between an asymptotic incoming state in the 
lead $m$, characterized by the collection $\alpha'_m$ of its quantum numbers, 
to an asymptotic outgoing state in the lead $n$, accordingly characterized 
by $\alpha_n$.

In Refs.~\onlinecite{Bar89,Noe93} the Landauer formalism was derived from
the Schr\"odinger equation in linear response theory, making use of an 
appropriate Kubo-Greenwood formula. We first remark that an inclusion of 
spin, interacting with the translational degrees of freedom via a Zeeman, 
spin-orbit, Rashba, or Dresselhaus coupling, into this method causes no 
problems. Although the current density is modified, its conservation in the 
form required for the Kubo-Greenwood expression of the conductivity to hold 
is indeed guaranteed. On then obtains for transmission (i.e. $m\neq n$),
\begin{equation}
\label{eq:gnm}
 g_{nm} = -\frac{e^{2}}{h} \int_{0}^{\infty} dE\, f_\beta'(E)
          \sum_{\alpha_n,\alpha_m'} \left| S^{nm}_{\alpha_n\alpha_m'}
          \right|^{2} \ ,
\end{equation}
and for reflection (i.e. $m=n$),
\begin{equation}
\label{eq:gnn}
 g_{nn} = \frac{e^{2}}{h} \int_{0}^{\infty} dE\, f_\beta'(E)
          \left( (2s+1)N_{n}-\sum_{\alpha_n,\alpha'_n}
          \left|S^{nn}_{\alpha_n\alpha'_n}\right|^{2} \right) \ .
\end{equation}
Here $N_n$ is the number of open channels in the lead $n$ (without spin
degeneracy) at energy $E$, and $f_\beta(E)$ denotes the Fermi distribution 
function at inverse temperature $\beta$. Of course, this requires the spin 
quantum number $s$ to be half-integer.

In a next step $S$-matrix elements have to be related to Green functions
$G(\mathbf{x},\mathbf{x}',E)$. These satisfy the equations
\begin{eqnarray}
\label{eq:Green1}
 & \left(\frac{1}{2m}\left(\hat{\mathbf{p}}-\frac{e}{c}
   \mathbf{A}(\hat{\mathbf{x}})\right)^{2}+\hat{\mathbf{s}}\cdot
   \mathbf{C}(\hat{\mathbf{x}},\hat{\mathbf{p}})-E\right)
   G\left(\mathbf{x},\mathbf{x}',E\right)\nonumber \\
 & =\delta (\mathbf{x}-\mathbf{x}') 
\end{eqnarray}
and
\begin{eqnarray}
\label{eq:Green2}
 & \left(\frac{1}{2m}\left(\hat{\mathbf{p}}'+\frac{e}{c}\mathbf{A}
   (\hat{\mathbf{x}}')\right)^{2}-E\right)G(\mathbf{x},\mathbf{x}',E)
   \nonumber \\
 &+\mathbf{C}^{\ast}(\hat{\mathbf{x}}',\hat{\mathbf{p}}')
   G(\mathbf{x},\mathbf{x}',E)\hat{\mathbf{s}} =\delta (\mathbf{x}-\mathbf{x}')
   \ .
\end{eqnarray}
The unusual form of the second equation is dictated by the fact that 
$G(\mathbf{x},\mathbf{x}',E)$ is a hermitian $(2s+1)\times (2s+1)$ matrix in 
spin space. In the following we will always choose advanced Green functions, 
fully characterized by Eqs.~(\ref{eq:Green1}) and (\ref{eq:Green2}) as well
as the condition that, asymptotically in the leads, they contain only 
outgoing contributions.

\begin{figure}
\begin{center}
\includegraphics[%
  bb=0bp 0bp 416bp 299bp,
  width=8cm]%
{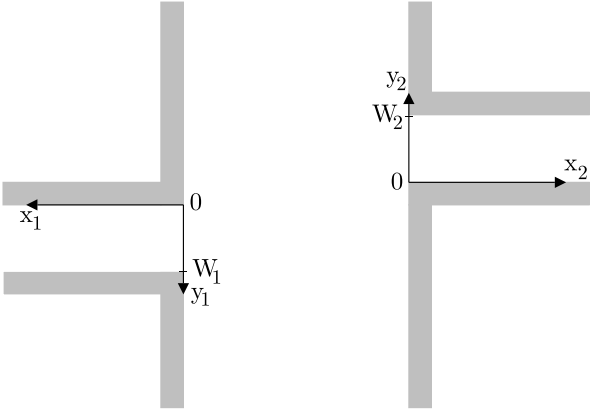}
\caption{Sketch of the geometry}\label{cap:fg4}
\end{center}
\end{figure}

As in the case without spin~\cite{Bar89,Noe93} one can then express the 
$S$-matrix elements in terms of the (advanced) Green function. Up to a global
phase factor, for $m\neq n$ this yields
\begin{eqnarray}
\label{eq:SGreen1}
 S^{nm}_{\alpha_n\alpha'_m} 
  & = & \frac{2\hbar^{2}}{im}\sqrt{\frac{k_{a_n}k_{a'_m}}{W_{m}W_{n}}}
        \int_{0}^{W_{n}}dy_{n}\int_{0}^{W_{m}}dy_{m}'\,\sin
        \left( \frac{a_n\pi y_{n}}{W_{n}} \right)  \nonumber \\
  &   & \times \sin \left( \frac{a'_m\pi y_{m}'}{W_{m}} \right) \, 
        G_{\sigma\sigma'}(\mathbf{x}_{n},\mathbf{x}_{m}',E) \ ,
\end{eqnarray}
and for $m=n$
\begin{eqnarray}
\label{eq:SGreen2}
 S^{nn}_{\alpha_n\alpha'_n} 
  & = & \frac{2\hbar^{2}}{im}\frac{\sqrt{k_{a_n}k_{a'_n}}}{W_{n}}
        \int_{0}^{W_{n}}dy_{n}\int_{0}^{W_{n}}dy_{n}'\,\sin
        \left( \frac{a_n\pi y_{n}}{W_{n}} \right)  \nonumber \\
  &   & \times\sin\left(\frac{a'_n\pi y_{n}'}{W_{n}}\right)\,G_{\sigma\sigma'}
        (\mathbf{x}_{n},\mathbf{x}'_{n},E)+\delta_{\alpha_n\alpha'_n}   \ .
\end{eqnarray}
Here we have introduced coordinates $\mathbf{x}_{n}=(x_n,y_n)$, where 
$x_n\geq0$ is a longitudinal, outward running coordinate in the lead $n$ and 
$0\leq y_n\leq W_n$ is the corresponding transversal coordinate (see also 
Figure~\ref{cap:fg4}). The transversal quantum number is $a_n=1,\dots,N_n$¸
with associated wave number $k_{a_n}=\sqrt{2mE/\hbar^2 - a_n^2\pi^2/W_n^2}$. 
The number $N_n$ of open transversal channels then is the largest integer 
$a_n$ that leaves the wave number real. Moreover, $\sigma=-s,\dots,s$ is a 
spin index such that altogether $\alpha_n = (E,a_n,\sigma)$.

We remark that in Eqs. (\ref{eq:SGreen1}) and (\ref{eq:SGreen2}) the points 
$\mathbf{x}_{n},\mathbf{x}_{m}'$ can be chosen anywhere in the respective 
leads. For later convenience we take them on the connection of the leads to 
the cavity, i.e., with $x_n =0 =x_m'$.
\subsection{Semiclassical Green function and transmission amplitudes}
In order to proceed further, one requires a semiclassical representation
for the Green function defined in Eqs.~(\ref{eq:Green1}) and (\ref{eq:Green2}).
In Ref.~\onlinecite{Bol99} this was achieved through an asymptotic expansion 
in powers of Planck's constant $\hbar$ for the quantum propagator generated 
by the Hamiltonian~(\ref{eq:hamilton}) which yielded, after a Fourier 
transformation, a respective semiclassical expansion for the Green function.
The range of validity of this procedure follows from the observation that,
since the spin operator $\hat{\mathbf{s}}$ is linear in $\hbar$, the energy
scale of the spin-orbit interaction term becomes small as compared to
the kinetic term in the limit $\hbar\to 0$. This condition is equivalent
to the spin-precession length being large compared to the Fermi wavelength.
In semiconductor heterostructures this requirement is usually fulfilled.

The semiclassical representation for the Green function obtained in 
Ref.~\onlinecite{Bol99} reads
\begin{equation}
\label{eq:sclgreen}
 G(\mathbf{x},\mathbf{x}',E) \sim \sum_{\gamma(\mathbf{x},\mathbf{x}')}
 A_{\gamma}(\mathbf{x},\mathbf{x}',E)\,\exp\left((i/\hbar)
 S_{\gamma}(\mathbf{x},\mathbf{x}',E)\right) \ ,
\end{equation}
as $\hbar\to 0$. The sum extends over all classical trajectories 
$\gamma(\mathbf{x},\mathbf{x}')$ generated by the classical Hamiltonian
\begin{equation}
\label{eq:classHam}
  H_0 (\mathbf{x},\mathbf{p}) = \frac{1}{2m} \left( \mathbf{p}- \frac{e}{c}
 \mathbf{A}\left(\mathbf{x}\right) \right)^{2}
\end{equation}
(plus reflections from hard walls) that run from $\mathbf{x}'$ to 
$\mathbf{x}$ at energy $E$. Choosing 
$(\mathbf{x},\mathbf{x}')=(\mathbf{x}_{n},\mathbf{x}_{m}')$ as in
(\ref{eq:SGreen1}), (\ref{eq:SGreen2}), the relevant trajectories are those
that enter the cavity at lead $m$ and leave through lead $n$. Moreover, 
$S_\gamma (\mathbf{x},\mathbf{x}',E)$ is the classical action of the 
trajectory, and the leading order of the amplitude 
$A_{\gamma}\left(\mathbf{x},\mathbf{x}',E\right)$ reads 
\begin{equation}
\label{eq:sclamplitude}
 A_{\gamma}(\mathbf{x},\mathbf{x}',E)=\frac{e^{-i\frac{\pi}{2}\nu_{\gamma}}}
 {i\hbar\sqrt{2\pi i\hbar}}\,\sqrt{C_{\gamma}}\,D_{\gamma}
 (\mathbf{x}',\mathbf{p}',t) \bigl( 1+O(\hbar) \bigr)\ . 
\end{equation}
Here $\nu_{\gamma}$ is a Maslov index of the trajectory $\gamma$, and  
\begin{equation}
\label{eq:Cdef}
 C_{\gamma} := \left|\det\begin{pmatrix} 
  \frac{\partial^{2}S_{\gamma}}{\partial\mathbf{x}\partial\mathbf{x}'} & 
  \frac{\partial^{2}S_{\gamma}}{\partial\mathbf{x}\partial E} \\
  \frac{\partial^{2}S_{\gamma}}{\partial\mathbf{x}'\partial E} & 
  \frac{\partial^{2}S_{\gamma}}{\partial E^{2}}
 \end{pmatrix}\right| \ .
\end{equation}
The contribution of the spin is, in leading semiclassical order, completely
contained in the spin-transport matrix $D_{\gamma}(\mathbf{x}',\mathbf{p}',t)$.
This is the spin-$s$ representation of the spin propagator
$d_{\gamma}(\mathbf{x}',\mathbf{p}',t)$, which is defined as a solution of 
the equation
\begin{equation}
\label{eq:spntrans}
 \frac{d}{dt} d_{\gamma}(\mathbf{x}',\mathbf{p}',t) + \frac{i}{2}\mathbf{C}
 \left(\mathbf{X}(t),\mathbf{P}(t)\right)\cdot\pmb{\sigma}\,
 d_{\gamma}(\mathbf{x}',\mathbf{p}',t)= 0
\end{equation}
with initial condition $d_{\gamma}(\mathbf{x}',\mathbf{p}',0)=1$. Here 
$(\mathbf{X}(t),\mathbf{P}(t))$ is the point in phase space of the 
classical trajectory $\gamma$ at time $t$. Its initial point at time $t=0$
is $(\mathbf{x}',\mathbf{p}')$. Moreover, $\pmb{\sigma}$ is the vector
of Pauli spin matrices. Therefore, $d_{\gamma}$ is an $\mathrm{SU}(2)$-matrix
that can be seen as a propagator for the spin along the classical
trajectory $\gamma$. 

Upon dividing the trajectory $\gamma$ into two pieces $\gamma_1$ and 
$\gamma_2$, such that $t=t_1+t_2$, the spin propagator is clearly 
multiplicative. Since $D_{\gamma}$ arises from a group representation it 
inherits this multiplicative property from the propagator, i.e., 
\begin{equation}
\label{eq:multprop}
 D_{\gamma}(\mathbf{x}',\mathbf{p}',t_{1}+t_{2}) = 
 D_{\gamma_2}(\mathbf{X}(t_{1}),\mathbf{P}(t_{1}),t_{2})\,
 D_{\gamma_1}(\mathbf{x}',\mathbf{p}',t_{1}) \  .
\end{equation}
This relation will be used extensively in Section~\ref{secIII}.

In order to obtain a semiclassical representation of transmission amplitudes
we insert the expression (\ref{eq:sclgreen}) into equation (\ref{eq:SGreen1}).
Then the integrals over $y$ and $y'$, respectively, are evaluated, 
asymptotically as $\hbar\to 0$, with the method of stationary phase. In 
this context we stress the following important observation: The number of 
accessible transversal states (including spin) in the $n$-th lead is 
$(2s+1)N_n=(2s+1)[\sqrt{2mE}W_n/(\pi\hbar)]$, where $[x]$ denotes the integer 
part of $x\in\mathbb{R}$. We choose the widths $W_n$ of the leads to 
formally shrink proportionally to $\hbar$ in this limit (compare also 
Ref.~\onlinecite{Heu06}) and hence set $W_n = \tilde{W}_n\hbar$, to the 
effect that the $\sin$-factors in Eqs.~(\ref{eq:SGreen1}) and 
(\ref{eq:SGreen2}) contribute rapidly oscillating phases. These have to 
be taken into account when determining stationary points of the total phases 
in the integrals. The condition of stationary phase hence imposes the 
following restrictions on the transversal momenta, 
\begin{equation}
\label{eq:13}
 p'_y = -\frac{\partial S_{\gamma}}{\partial y'_{m}} = 
        \pm\frac{a'_{m}\pi}{\tilde{W}_{m}}
\end{equation}
and
\begin{equation}
\label{eq:14}
 p_y = \frac{\partial S_{\gamma}}{\partial y_{n}} =
       \mp\frac{a_{n}\pi}{\tilde{W}_{n}} \ ,
\end{equation}
upon entry and exit, respectively, of the trajectories. If the points of 
entry and exit are free of magnetic fields, and thus 
$\mathbf{p}=m\dot{\mathbf{x}}$ at these points, one can characterize the 
trajectories in terms of the angles $\theta$ and $\theta'$, under which they 
enter and leave the cavity with respect to the longitudinal directions of 
the leads (see also Figure~\ref{cap:fg4a}). These angles are related to 
the transversal momenta (\ref{eq:13}) and (\ref{eq:14}) through 
$\sin\theta = p_y/\sqrt{2mE}$ and $\sin\theta' = p'_y/\sqrt{2mE}$.
If one wished to keep the widths of the openings fixed, however, the
method of stationary phase would enforce the conditions $p_y' =0= p_y$ upon
the trajectories, thus leading to different semiclassical expressions than
the ones we use henceforth. 

Collecting now all terms that emerge in the stationary phase calculation 
finally leads to the following leading semiclassical contribution to the
$S$-matrix elements,
\begin{equation}
\label{eq:12}
 S^{nm}_{\alpha_n\alpha'_m} \sim \sum_{\gamma(\theta,\theta')} 
 B_{\gamma(\theta,\theta')}
 \,D^{\sigma\sigma'}_{\gamma(\theta,\theta')}\,\exp\left(\left(i/\hbar\right)
 S_{\gamma(\theta,\theta')}\right) \ ,
\end{equation}
where the sum extends over all trajectories that run from lead $m$ through 
the cavity to lead $n$ and are characterized by the conditions (\ref{eq:13}), 
(\ref{eq:14}), expressed in terms of the angles of entry and exit. The 
explicit form of the factor $B_{\gamma(\theta,\theta')}$ is the same as if 
there were no spin present\cite{Bar89}, 
\begin{equation}
\begin{split}
\label{eq:15}
 B_{\gamma(\theta,\theta')} = 
 &\sqrt{\frac{i\pi\hbar}{2W_{m}W_{n}}}\,\frac{\mathrm{sgn}(\pm a_{m}')
    \mathrm{sgn}(\pm a_{n})}{|\cos\theta\cos\theta' 
    M_{\gamma(\theta,\theta')}^{21}|^{1/2}} \\ 
 &\times\exp\left( i\pi\left(\frac{\pm a_{m}' y'_{m}}{W_{m}}+
  \frac{\pm a_{n}y_{n}}{W_{n}}-\frac{1}{2}\mu_{\gamma(\theta,\theta')}
  \right)\right) \ .
\end{split}
\end{equation}
Here $M_{\gamma(\theta,\theta')}^{21}$ is an element of the monodromy matrix
of $\gamma(\theta,\theta')$ that arises from the matrix appearing in 
(\ref{eq:Cdef}) by a restriction to the phase space directions transversal 
to the trajectory. Furthermore, $\mu_{\gamma(\theta,\theta')}$ is a modified 
Maslov index that contains the index $\nu_{\gamma(\theta,\theta')}$ from 
Eq.~(\ref{eq:sclamplitude}) and additional phases resulting from the 
stationary phase calculation of the integrals over $y_n$ and $y_m'$. 

The above result (\ref{eq:12}) primarily refers to transmission amplitudes
($n\neq m$), but can be carried over to the case of reflection ($n=m$). The 
reason for this is that the additional term $\delta_{\alpha_n\alpha'_n}$ in 
(\ref{eq:SGreen2}) is canceled by the contribution of direct trajectories
in the opening of the lead $n$ that never enter the cavity \cite{Bar93}.

The ultimate goal being a semiclassical calculation of the conductance
coefficients (\ref{eq:gnm}) and (\ref{eq:gnn}), one therefore requires
the evaluation of double sums
\begin{equation}
\begin{split}
\label{eq:15.5}
 \left| S^{nm}_{\alpha_n\alpha_m'} \right|^{2} \sim 
 \sum_{\gamma(\theta,\theta')} \sum_{\gamma'(\theta,\theta')} 
 &B_{\gamma}B_{\gamma'}^\ast \,D^{\sigma\sigma'}_{\gamma}
  {D^{\sigma\sigma'}_{\gamma'}}^\ast \\
 &\times\exp\left(\left(i/\hbar\right)\left(S_{\gamma}-S_{\gamma'}
  \right)\right) 
\end{split}
\end{equation}
over classical trajectories. This will be the task for the rest of this paper.

To simplify the calculations, from now on we restrict our attention to
the case of two leads. With an incoming wave in the lead $m=1$ we are thus 
dealing with the transmission coefficient $g_{21}$ and the reflection
coefficient $g_{11}$. To this end we will determine the transmission matrix 
$S^{21}$ and the reflection matrix $S^{11}$, leading to the transmission
and reflection coefficient
\begin{equation}
\label{transrefl}
 {\cal T} = \sum_{\alpha_2,\alpha_1'} \left| S^{21}_{\alpha_2\alpha_1'}
 \right|^{2} \ ,
 \quad {\cal R} = \sum_{\alpha_2,\alpha_2'} \left| S^{22}_{\alpha_2\alpha_2'}
 \right|^{2}\ ,
\end{equation}
respectively. Hence, at zero temperature the current through lead 2 is
\begin{equation}
\label{current}
 I_2 = \frac{e^2}{h}\,\left( {\cal T}\,V_1 +\left(  {\cal R}\,-\left( 2s+1
       \right) N_2 \right) V_2 \right) \ ,
\end{equation}
where ${\cal T}$ and ${\cal R}$ are taken at the Fermi energy $E_F$.
Together with the condition $g_{21}+g_{22}=0$, expressing that equal 
voltages at both leads produce no current, this yields the relation
\begin{equation}
\label{cur}
 I_2=\frac{e^2}{h}{\cal T}\,\left( V_1-V_2 \right) \ . 
\end{equation}
\section{Semiclassical calculation of conductivity coefficients}
\label{secIII}
The calculation of the double sum (\ref{eq:15.5}) over classical trajectories 
requires input from dynamical properties of the associated classical
system. With spin-orbit interactions present, one therefore first has
to identify an appropriate classical system. Moreover, ergodic properties 
of the classical system imply necessary ingredients for the further
calculation. The diagonal contribution to the double sum is evaluated with
a sum rule \cite{Sie02,Zai05}, whereas the non-diagonal terms are
evaluated following the Sieber-Richter method \cite{Ric01,Sie02,Zai05,Heu06}.
\subsection{Classical spin-orbit dynamics}
The classical dynamics that enter the semiclassical representation
(\ref{eq:sclgreen}) consist of two parts\cite{Bol99}: the motion of the 
point particle generated by the Hamiltonian (\ref{eq:classHam}), including 
elastic reflections from hard walls, and the spin that is driven by this 
motion according to (\ref{eq:spntrans}). These contributions can be combined 
into a single dynamics on a spin-orbit phase space\cite{Bol99a}. The relevant 
classical trajectory is $(\mathbf{X}(t),\mathbf{P}(t),g(t))$, with initial 
condition $(\mathbf{x}',\mathbf{p}',g')$ at $t=0$. Here $g\in\mathrm{SU}(2)$ 
and $g(t)=d_{\gamma}(\mathbf{x}',\mathbf{p}',t)g$ provides the spin part of 
the combined motion. We remark that this description of spin appears quantum 
mechanical. However, by passing to expectation values of the spin operator 
$d_{\gamma}^\dagger\frac{1}{2}\pmb{\sigma}d_{\gamma}$ in normalized spin
states $\chi$ (Heisenberg picture), the spin variable becomes a unit vector
$\langle\chi,d_{\gamma}^\dagger\frac{1}{2}\pmb{\sigma}d_{\gamma}\chi\rangle$. 
Hence the spin part of the combined phase space is a unit sphere. The two 
views of the spin motion, either on $\mathrm{SU}(2)$ or on a unit sphere, 
are in fact equivalent\cite{Bol99}. In both cases we will therefore speak of
classical spin-orbit dynamics.
 
Ergodicity is a concept developed for closed systems. It can, however, 
be suitably extended to open systems of the kind under consideration here.
To this end one divides the configuration space $Q$ of the device into a
closed part $Q_c$, consisting of the cavity with the leads truncated and 
the openings closed, plus the infinite leads. From now on we suppose the
shape of the closed part to form a chaotic billiard, ensuring ergodicity
of the motion inside the cavity. Then $\rho(t)$ is the probability for a
typical trajectory to stay within the cavity at least up to time $t$.
For large times,
\begin{equation}
\label{eq:17.3}
 \rho(t) \sim \exp\left(-t/\tau\right) \ ,\quad t\to\infty \ ,
\end{equation}
with inverse dwell time
\begin{equation}
\label{eq:17.4}
 \frac{1}{\tau}=\frac{\hbar}{mA}(N_1 +N_2 ) \ ,
\end{equation}
in which $A$ denotes the area of the closed part $Q_c$. For the associated 
part of phase space we also introduce the volume
\begin{equation}
\label{eq:17.1}
 \Sigma(E) = \int_{Q_c}d^{2}x\int_{\mathbb{R}^{2}}d^{2}p\,
 \delta\left(E-H_0(\mathbf{x},\mathbf{p})\right) = 2\pi mA
\end{equation}
of the energy shell. This expression has no integration over the spin
part, since the Hamiltonian is independent thereof, and an integration
over $\mathrm{SU}(2)$ with respect to Haar measure $dg$ yields one.

For the open system the concept of ergodicity has to be modifed in that
the possibility of a trajectory to leave the cavity must be taken into 
account. When the motion inside the cavity is ergodic this leads to the
following relation between phase-space averages and time averages over typical 
spin-orbit trajectories,
\begin{eqnarray}
\label{eq:19}
 & \left\langle\int_{0}^{T}dt\, f\left(\mathbf{X}(t),\mathbf{P}(t),g(t)\right)
   \right\rangle \sim \frac{1}{\Sigma\left(E\right)}\int_{0}^{T}dt\,
   \rho(t)\times\nonumber \\
 & \int_{Q_c}d^{2}x\int_{\mathbb{R}^{2}}d^{2}p\int_{SU(2)}dg 
   f(\mathbf{x},\mathbf{p},g)\,\delta\left(E-H_0(\mathbf{x},\mathbf{p})
   \right) \ ,
\end{eqnarray}
as $T\to\infty$. Here $f$ is an arbitrary function on the combined phase
space, and $\langle\dots\rangle$ denotes an average over initial conditions.
This relation, which properly reflects the chaotic nature of the combined 
classical spin-orbit motion, provides the basis for the further use of
dynamical properties in the calculation of the sum (\ref{eq:15.5}) over
classical trajectories.

The stronger mixing property, which we also assume to hold henceforth,
means that correlations of two observables $f$ and $h$ decay, i.e., 
\begin{widetext}
\begin{eqnarray}
\label{eq:20}
 &  \lim_{t\rightarrow\infty}\int_{Q_c}d^{2}x\int_{\mathbb{R}^{2}}d^{2}p
 \int_{SU(2)} dg\ h\left(\mathbf{X}(t),\mathbf{P}(t),g(t) \right)
 f\left(\mathbf{x},\mathbf{p},g\right)\delta\left(E-H_0(\mathbf{x},\mathbf{p})
 \right)\nonumber \\
 &  =\frac{1}{\Sigma(E)}\int_{Q_c}d^{2}x\int_{\mathbb{R}^{2}}d^{2}p
    \int_{SU(2)}dg\ h(\mathbf{x},\mathbf{p},g)\delta\left(E-H_0(\mathbf{x},
    \mathbf{p})\right)\ \int_{Q_c}d^{2}x'\int_{\mathbb{R}^{2}}d^{2}p'
    \int_{SU(2)}dg'f(\mathbf{x}',\mathbf{p}',g')\delta\left(E-H_0(\mathbf{x}',
    \mathbf{p}')\right) \ .
\end{eqnarray}
\end{widetext}
\subsection{Transmission and reflection coefficients in the 
configuration-space approach}
In a first step we calculate the leading semiclassical contribution to 
transmission and reflection coefficients from equation (\ref{eq:15.5}), 
averaged over a small energy window, by using the configuration-space 
approach. Such a calculation has been performed previously\cite{Zai05}, 
however, with a sum rule that only takes the particle motion into account. 
The spin contribution was built in subsequently, assuming that traces of
products of spin-transport matrices can be replaced by certain averages.
Here we reproduce the result obtained in Ref.~\onlinecite{Zai05} by using a 
sum rule for the complete spin-orbit dynamics that follows from (\ref{eq:19}). 
Thus we base the assumptions made in Ref.~\onlinecite{Zai05} on a firm 
dynamical ground.

As $\hbar\to 0$ the terms in the double sum~(\ref{eq:15.5}) are highly 
oscillatory, except for contributions with $S_{\gamma}=S_{\gamma'}$. 
Generically, if no symmetries are present, this only occurs for the diagonal 
$\gamma'=\gamma$. In the event that time-reversal invariance is not broken,
however, the time-reversed trajectory $\gamma^{-1}$ has the same action
as $\gamma$. Of course, $\gamma^{-1}$ is only among the trajectories to
be summed over in the case of reflection ($n=1=m$) when, moreover, 
$\theta=\theta'$; i.e., only for $S^{11}_{\alpha_1\alpha_1'}$ with
$a_1=a_1'$. All further terms are oscillatory, with a decreasing importance 
of their contribution, after averaging over an energy window, when the
action differences increase. Below we calculate the two leading 
contributions to the quantity
\begin{equation}
\label{eq:19.1}
 \sum_{\sigma,\sigma'=-s}^{s}\left| S^{nm}_{\alpha_n\alpha_m'} \right|^{2}
 \sim \sum_{\gamma,\gamma'} B_{\gamma}B_{\gamma'}^\ast \,
 \mathrm{Tr}\,(D_{\gamma}D_{\gamma'}^\dagger)
  \exp\left(\left(i/\hbar\right)\left(S_{\gamma}-S_{\gamma'}\right)\right)
\end{equation}
for systems with time-reversal invariance: (i) the diagonal contribution in 
which the sum over $\gamma'$ is restricted to $\gamma' =\gamma$ (for 
transmission) or $\gamma' =\gamma^{\pm 1}$ (for reflection), and (ii) the 
one-loop contribution in which the sums over $\gamma$ and $\gamma'$ are 
confined to so-called Sieber-Richter pairs (see also Ref.~\onlinecite{Zai05}). 

Due to the unitarity of the spin-transport matrices, in the diagonal case 
terms with $\gamma'=\gamma$ yield a spin contribution of 
$\mathrm{Tr}\,(D_{\gamma}D_{\gamma}^\dagger) =2s+1$. Thus, the diagonal
contribution to (\ref{eq:19.1}) can immediately be obtained from the 
respective result without spin\cite{Bar93,Sie02},
\begin{equation}
\label{eq:24}
 \Bigl\langle\sum_{\sigma,\sigma'=-s}^{s}\left| S^{nm}_{\alpha_n\alpha_m'} 
 \right|^{2}_{\mathrm{diag}}\Bigr\rangle_{\Delta E} 
 \sim \frac{2s+1}{N_1 +N_2} \ . 
\end{equation}
In the case of reflection ($n=1=m$) with $a_1=a_1'$ an additional diagonal
contribution arises from the terms with $\gamma' =\gamma^{-1}$, if
time-reversal invariance is unbroken. Its spin contribution is
$\mathrm{Tr}\,(D_{\gamma}D_{\gamma^{-1}}^\dagger)=\mathrm{Tr}\,(D_{\gamma}^2)$.
One hence requires a suitable sum rule that incorporates the combined 
classical spin-orbit motion. For this purpose we choose the function 
\begin{equation}
\begin{split}
\label{eq:testfct}
 f&\left(\mathbf{X}(t),\mathbf{P}(t),g(t)\right) \\
  &=\frac{1}{m}\delta(\vartheta(t)-\theta)\,\delta(x(t))\,
    \bigl( \Theta(y(t)) - \Theta(y(t)-W_1) \bigr) \\
  &\quad\times \,\mathrm{Tr}\,\bigl( \pi_s (g(t) g(0)^{-1}) \bigr)^2
\end{split}
\end{equation}
in (\ref{eq:19}). Here $\pi_s(g)$ denotes the spin-$s$ representation of
$g\in\mathrm{SU}(2)$, $\vartheta$ is the angular variable in planar polar
coordinates for $\mathbf{p}$ and $\Theta(y)$ is a Heavyside step function.
An evaluation of (\ref{eq:19}) with the function (\ref{eq:testfct}) then leads 
to the sum rule (as $T\rightarrow\infty)$
\begin{equation}
\label{eq:23}
 \sum_{\gamma,T_{\gamma}\leq T} \left| B_{\gamma}\right|^{2}\,
 \mathrm{Tr}\,(D_{\gamma}^2)  
 \sim\frac{\pi}{2\tilde{W}_1}\frac{\left(-1\right)^{2s}}{2\pi mA}
 \int_{0}^{T}dt\,\rho(t) \ .
\end{equation}
After an average over a small window in energy this, together with 
(\ref{eq:24}), finally yields the semiclassical result
\begin{equation}
\label{eq:24a}
 \Bigl\langle\sum_{\sigma,\sigma'=-s}^{s}\left| S^{11}_{\alpha_1\alpha_1'} 
 \right|^{2}_{\mathrm{diag}}\Bigr\rangle_{\Delta E} \sim 
 \frac{2s+1+(-1)^{2s}\delta_{a_1 a_1'}}{N_1 +N_2}  
\end{equation}
for the diagonal contribution to (\ref{eq:19.1}). For $s=1/2$ the right-hand
side is $1/(N_1 +N_2)$.

\begin{figure}

\begin{center}
\includegraphics[%
  bb=0bp 0bp 416bp 299bp,
  width=8cm]{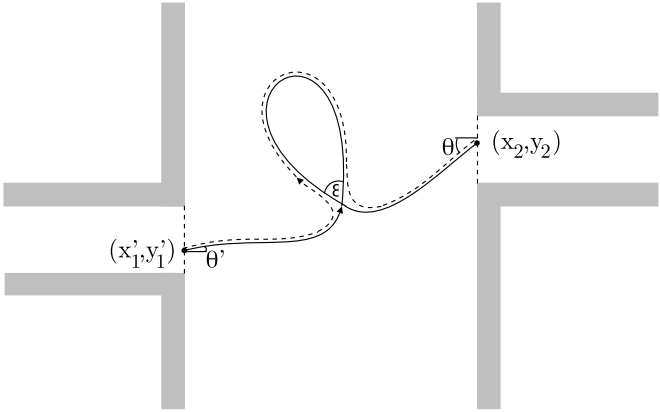}
\caption{A Sieber-Richter pair of trajectories}\label{cap:fg4a}
\end{center}
\end{figure}

Sieber-Richter pairs of trajectories are characterized by the fact that
one trajectory possesses a self-crossing with a small crossing angle 
$\epsilon$, thus forming a loop. The partner trajectory then looks like
the former one cut open at the self-crossing, but with the loop direction
reversed and then glued together, such that the self-crossing is replaced by 
an almost-crossing, see Figure~\ref{cap:fg4a}. In principle, the trajectories 
in such pairs can have an 
arbitrary number of self-crossings, but the magnitude of their contributions 
to (\ref{eq:19.1}) decreases with increasing numbers of places in which 
self-crossings are paired with almost-crossings. The most important 
('one-loop') contribution comes from pairs which differ in one crossing. In 
order to calculate the one-loop contribution one requires the distribution of 
the crossing angles $\epsilon$ for pairs of trajectories with loops of 
duration $T$,
\begin{equation}
\label{eq:25}
 P_{S}(\epsilon,T) = \frac{1}{\Sigma(E)}\int_{Q_c}d^{2}x'\int_{\mathbb{R}^{2}}
 d^{2}p'\int_{T_{min}(\epsilon)}^{T}dt_{l} \, 
 p_{S}(\epsilon,T,t_{l}) \ .
\end{equation}
Here $p_{S}(\epsilon,T,t_{l})$ is a density of crossing angles defined as
\begin{eqnarray}
\label{eq:26}
 p_{S}(\epsilon,T,t_{l}) 
 &=&
\int_{0}^{T-t_{l}}dt_{s}\,|J|\,\delta(E-H_0(\mathbf{P}(t_{s})))
    \nonumber \\
& &\times \mathrm{Tr}\left(\pi_{s}\left[g(t)(g(0))^{-1}\right]\right)^{2} 
   \delta(\epsilon-\kappa(t_{s},t_{l})) \nonumber \\
& &\times\delta(\mathbf{X}(t_{s})-
   \mathbf{X}(t_{s}+t_{l})) \ ,
\end{eqnarray}
where $\kappa(t_{s},t_{l})$ denotes the angle between the velocities 
$\mathbf{v}(t_{s})$ and $\mathbf{v}(t_{s}+t_l)$. Given a crossing angle 
$\epsilon$, the minimal duration for a loop to close is $T_{min}(\epsilon)$.
In chaotic systems this quantity behaves like 
$T_{min}(\epsilon)=O(\log\epsilon)$ as $\epsilon\to 0$\cite{Ric01}. 
Furthermore,  
\begin{equation}
\begin{split}
\label{eq:28}
 |J| &= | \mathbf{v}(t_{s}) \times \mathbf{v}(t_{s}+t_l) | \\
     &= | \mathbf{v}(t_{s}) |\, |\mathbf{v}(t_{s}+t_l) | \,
        \sin\kappa(t_{s},t_{l})
\end{split}
\end{equation}
is a Jacobian, and $t_s$, $t_l$ denote the time along the trajectory up to
the starting point of the loop and along the loop, respectively. 

Assuming that the classical spin-orbit dynamics are not only ergodic, but 
also mixing, the distribution (\ref{eq:25}) can be calculated further. It 
can be identified as the left-hand side of an appropriate relation of the type
(\ref{eq:20}). The right-hand side then yields, as $\epsilon\to 0$,
\begin{equation}
\label{eq:29}
 P_{S}(\epsilon,T) \sim \frac{(-1)^{2s}}{\pi A}\,\frac{2E}{m}\,\sin\epsilon
 \left(\frac{T^{2}}{2}-TT_{min}\left(\epsilon\right)+\frac{T_{min}^{2}
 \left(\epsilon\right)}{2}\right) \ .
\end{equation}
This expression differs from the respective one without spin that was obtained 
in Ref.~\onlinecite{Ric01} only by a factor $(-1)^{2s}$, i.e., a sign in the
case $s=1/2$. With this information at hand the one-loop contribution can
be calculated as in the case without spin\cite{Sie02}, finally yielding
\begin{equation}
\label{eq:29.5}
 \Bigl\langle\sum_{\sigma,\sigma'=-s}^{s}\left|S^{21}_{\alpha_1\alpha_2'}
 \right|_{\mathrm{1-loop}}^{2}\Bigr\rangle_{\Delta E} \sim -\frac{(-1)^{2s}}
 {\left(N_{1}+N_{2}\right)^{2}} \ .
\end{equation}
This is in accordance with what has been obtained in Ref.~\onlinecite{Zai05}.
\subsection{Transmission coefficients in the phase-space approach}
Higher orders in the 'loop-expansion' described above have been calculated
previously for spectral form factors\cite{Heu05} as well as for conductance
coefficients for systems without spin contributions\cite{Heus06}. The
approach taken in these papers utilizes trajectories in classical phase
space and identifies the pairs of self-crossings/almost-crossings in
configuration space as pairs of trajectories with almost-crossings in phase 
space, which differ in the way they are connected at the (almost) crossings.
This point of view opens the possibility for a classification of the 
trajectory pairs in terms of their encounters\cite{Heu05}. Here we follow 
this phase-space approach and amend the previous result\cite{Heus06} with 
the contribution of the spin-orbit coupling. 

To be more precise, we consider trajectories that possess close 
self-encounters (in phase space), in which two or more short stretches of 
the trajectory are almost identical, possibly up to time reversal. These 
stretches are connected by long parts of the trajectory, which we call 
loops. We then form pairs  $(\gamma,\gamma')$ of such trajectories in which
$\gamma$ and $\gamma'$ are almost identical (up to time reversal) along the 
loops, but differ from each other in the way the loops are connected in the 
encounter region. In order to quantify these encounters, we introduce a 
vector $\vec{v}$, whose $l$-th component, $v_{l}$, denotes the number of 
encounters with $l$ stretches. Hence the total number of encounters is 
$V=\sum_{l\geq 2} v_l$, with a total of $L=\sum_{l\geq 2}lv_{l}$ stretches 
involved. In general, however, given a vector $\vec{v}$, there will be
$N\left(\vec{v}\right)\geq 1$ different trajectory pairs associated with it. 
These may, e.g., differ in the order the loops connect the encounters, or in 
the relative directions, in which the encounter-stretches are traversed. 

To reveal the phase-space structure of trajectory pairs and to compute
their contributions to (\ref{eq:15.5}) one introduces Poincar\'{e} sections,
which cut the trajectories into pieces. In order to adapt this cutting
to the sequence of encounters and loops one chooses a Poincar\'{e} section 
in every of the $V$ given encounter regions. We then denote by 
$t_{\alpha,j}'$, $j=1,\dots,l_\alpha$, $\alpha=1,\dots,V$ the times at 
which the encounter stretches pierce this section, and by $t_{enc}^\alpha$, 
the duration of the encounters. To this cutting of the trajectories 
corresponds the splitting 
\begin{equation}
\label{eq:30}
 D_{\gamma} = D_{L+1}D_{L}...D_{1} 
\end{equation}
of the spin-transport matrices which, with an obvious notation, follows from 
the composition rule (\ref{eq:multprop}). The spin transport along the
partner trajectory then reads
\begin{equation}
\label{eq:31}
 D_{\gamma'} \approx D_{L+1}D_{k_{L}}^{\eta_{L}}...D_{k_{2}}^{\eta_{2}}D_{1} 
 \ .
\end{equation}
Here $\eta_j=\pm 1$, depending on the relative orientation of the trajectory
between the $j-1$-st and the $j$-th cutting of $\gamma$ and $\gamma'$,
respectively, through the Poincar\'{e} section. We notice that at this
point time-reversal invariance enters crucially. Moreover, the indices $k_{j}$ 
take care of the fact that in $\gamma$ and $\gamma'$ the loops may be 
traversed in different successions. Thus the spin-dependent weights in
(\ref{eq:19.1}) for each pair of trajectories are approximately given by 
\begin{equation}
\label{eq:31.1}
 \mathrm{Tr}\bigl( D_{\gamma}D_{\gamma'}^\dagger \bigr) \approx 
 \mathrm{Tr}\bigl( D_{L}...D_{2}D_{k_{2}}^{\dagger\eta_{2}}\dots
 D_{k_{L}}^{\dagger\eta_{L}} \bigr) \ .
\end{equation}

The calculation of transmission amplitudes performed in 
Ref.~\onlinecite{Heus06} has now to be modified in that the expressions 
(\ref{eq:31.1}) must be included. To this end we recall the strategy devised
in Refs.~\onlinecite{Heu05,Heus06}: For each encounter one introduces
coordinates on the Poincar\'e section adapted to the piercing by the 
trajectories and the linear stability of the dynamics. In encounter $\alpha$ 
the coordinates $(s^\alpha_j,u^\alpha_j)$, $j=1,\dots,l_\alpha -1$, describe
the separation of the $j+1$-st piercing from the $j$-th one along the stable
and unstable manifolds, respectively, of the latter. The total of $L-V$
stable and unstable coordinates are then collected in the vectors
$(\mathbf{s},\mathbf{u})$. In these coordinates action differences of 
partner trajectories (approximately) read as
\begin{equation}
\label{eq:41.5}
 \Delta S=S_\gamma -S_{\gamma'}\approx\sum_{\alpha,j} s^\alpha_j 
 u^\alpha_j \ .
\end{equation}
Moreover, the requirement that encounters be close can then be expressed in 
terms of the condition $|s^\alpha_j|,|u^\alpha_j|\leq c$ with some constant 
$c$, which yields the duration of an encounter
\begin{equation}
\label{eq:41.6}
 t_{enc}^\alpha\sim\frac{1}{\lambda}\ln\frac{c^2}{\max_i\left\lbrace\left|s_i 
 \right|\right\rbrace\max_j\left\lbrace\left|u_j \right|\right\rbrace }, 
 \quad t_{enc}^\alpha\rightarrow\infty\ .
\end{equation}
One then introduces a density 
$w_T^{\mathrm{spin}}(\mathbf{s},\mathbf{u})$ of encounters, weighted with 
the spin contribution, for trajectories of duration $T$ with a given encounter 
structure specified by the vector $\vec{v}$. In analogy to the case without 
spin\cite{Heu06} this leads to the following approximation,
\begin{equation}
\begin{split}
\label{eq:42}
 \Bigl\langle \sum_{\gamma}\sum_{\vec{v}}N(\vec{v})\int_{-c}^c\dots
 \int_{-c}^c 
 &d^{L-V}u\,d^{L-V}s\,\exp\left((i/\hbar)\Delta S\right)  \\
 &\times w_T^{\mathrm{spin}}(\mathbf{s},\mathbf{u}) \,|B_{\gamma}|^2 
  \Bigr\rangle _{\Delta E} \ ,
\end{split}
\end{equation}
to the quantity
\begin{equation}
\label{eq:loopstart}
 T_{a_2 a_1'}^{\mathrm{nd}} := \Bigl\langle\sum_{\sigma,\sigma'=-s}^{s}
 \left| S^{21}_{\alpha_2\alpha_1'} \right|^{2} 
 \Bigr\rangle_{\Delta E} - \frac{2s+1}{N_{1}+N_{2}} \ .
\end{equation}
After summing over all possible values of $a_2, a_1'$, this yields the 
non-diagonal contribution to the 
energy-averaged transmission amplitude ${\cal T}$, compare 
(\ref{transrefl}), (\ref{eq:24}).

The essential point now is to calculate the density 
$w_T^{\mathrm{spin}}(\mathbf{s},\mathbf{u})$. In the case without spin-orbit 
interaction the corresponding expression $w_T(\mathbf{s},\mathbf{u})$ was 
defined in Ref.~\onlinecite{Heu05} as a density of phase-space separations 
$\mathbf{s}$ and $\mathbf{u}$ similar to the density $P( \epsilon,T)$ with 
respect to $\epsilon$ in the configuration-space approach. It was given as
\begin{eqnarray}
\label{eq:w_T}
 w_{T}\left(\mathbf{s},\mathbf{u}\right)
 &=& \frac{1}{\Sigma(E)}\int_{Q_c}d^{2}x'\int_{\mathbb{R}^{2}}d^{2}p'
     \delta\left(E-{H_0}(\mathbf{x}',\mathbf{p}')\right)\nonumber \\
 & & \times\int_{0}^{\infty}\prod_{j=1}^{L}dt_{j} \ \Theta
     \left(T-\sum_{\alpha=1}^{V}l_{\alpha}t_{enc}^{\alpha}-\sum_{j=1}^{L}
     t_{j}\right)\nonumber \\
 & & \times \prod_{\alpha=1}^{V}\frac{1}{t_{enc}^{\alpha}} \left( 
     \prod_{j=2}^{l_\alpha}\delta\left(\left(\mathbf{X}(t'_{\alpha j}),
     \mathbf{P}(t'_{\alpha j})\right)-z_{\alpha j}\right)\right) 
     \ .\nonumber \\ 
\end{eqnarray}
The average in the first line is over all possible initial points of the 
trajectory. In the second line the integration extends over all loop 
durations $t_j$; their lengths are constrained by the theta function. In 
order to prevent over-counting\cite{Heu05}, the product of all encounter 
durations $t_{enc}^\alpha$ is divided out. The last product guarantees that 
the position of the orbit at times when it pierces through the sections 
are fixed as $z_{\alpha j}$. This denotes the first point of the orbit in 
which it pierces through a certain section plus the separation thereof as
specified by the coordinates $\mathbf s$ and $\mathbf u$. From 
Eq.~(\ref{eq:w_T}) one obtains $w_T^{\mathrm{spin}}(\mathbf{s},\mathbf{u})$ 
by including $\mathrm{Tr}( D_\gamma D_{\gamma'}^\dagger)$ under the integral. 
Using that the durations of encounters are semiclassically large, 
compare~(\ref{eq:41.6}), the result can be obtained in analogy to 
(\ref{eq:26}) by employing (\ref{eq:20}). The right-hand side then yields 
\begin{eqnarray}
\label{eq:34}
 w_{T}^{\mathrm{spin}}(\mathbf{s},\mathbf{u})
 &\approx& \frac{\left(T-\sum_{\alpha=1}^{V}l_{\alpha}t_{enc}^{\alpha}
           \right)^L}{\Sigma(E)^{L-V}\prod_{\alpha=1}^V t_{enc}^{\alpha}
           L!} \, M_{\gamma\gamma'} \ ,
\end{eqnarray}
i.e. a factorization into the spin-independent part identical to 
$w_{T}(\mathbf{s},\mathbf{u})$ and a spin contribution
\begin{eqnarray}
\label{eq:35}
 M_{\gamma\gamma'} 
 &:=& \int_{SU(2)}\ldots\int_{SU(2)}dg_{L}\ldots dg_{2}\nonumber \\
 &  & \times \mathrm{Tr}\left(\pi_{s}\left(g_{L}\ldots g_{2}  
      g_{k_{2}}^{\eta_{2}\dagger}\ldots g_{k_{L}}^{\eta_{L}\dagger}
      \right)\right) \ .
\end{eqnarray}
In order to calculate (\ref{eq:35}) we follow the method developed in
Refs.~\onlinecite{Bol03,Bol05} for the spectral form factor 
of quantum graphs with spin-orbit interaction. In analogy to 
Theorem~6.1 of Ref.~\onlinecite{Bol05} we find in the present context that
\begin{equation}
\label{eq:36}
 M_{\gamma\gamma'} = (2s+1) \left(\frac{(-1)^{2s}}{2s+1}\right)^{L-V} \ .
\end{equation}
This will be proven in the appendix. We stress that this spin contribution,
apart from the spin quantum number, only depends on $L-V$. 

The quantity (\ref{eq:loopstart}) can now be calculated in analogy to the 
case without spin~\cite{Heus06}. Starting from equation~(\ref{eq:42}),
one employs the expressions for $\Delta S$ from (\ref{eq:41.5}) and for
$w_{T}^{\mathrm{spin}}(\mathbf{s},\mathbf{u})$, the sum rule from 
Ref.~\onlinecite{Sie02} and the survival probability $\rho\left(t\right)$,
modified by replacing $t$ with 
$(t-\sum_{\alpha=1}^{V} (l_\alpha-1) t_{enc}^{\alpha})$ as in 
Ref.~\onlinecite{Heus06}. This yields 
\begin{widetext}
\begin{eqnarray}
\label{eq:43}
 T_{a_2,a_1'}^{nd} &\approx&
   \left\langle\frac{(2s+1)\hbar}{mA}\sum_{\vec{v}}N(\vec{v})\left(
   \prod_{i=1}^{L+1}\int_{0}^{\infty}dt_{i}\exp\left(-\frac{t_{i}}{\tau}
   \right)\right)\int_{-c}^{c}\ldots\int_{-c}^{c}\frac{d^{L-V}u d^{L-V}s}
   {(\Sigma(E))^{L-V}}\prod_{\alpha=1}^{V}\frac{\exp\left(-\frac{
   t_{enc}^{\alpha}}{\tau}+\frac{i}{\hbar}\Delta S\right)}{t_{enc}^{\alpha}}
   \right\rangle_{\Delta E}\left(\frac{(-1)^{2s}}{2s+1}\right)^{L-V}
   \nonumber \\
 &\approx& \frac{(2s+1)}{N_{1}+N_{2}}\sum_{k=1}^{\infty}\left(\frac{1}{N_{1}
   +N_{2}}\right)^{k}\left(\frac{(-1)^{2s}}{2s+1}\right)^{k}
   \sum_{\vec{v},L-V=k}(-1)^{V} N(\vec{v})\ .
\end{eqnarray}
\end{widetext}
The integrals over $\mathbf{s}$ and $\mathbf{u}$ were calculated
in Ref.~\onlinecite{Heus06}, and the sum over $\vec{v}$ can be carried out 
with the recursion formula~\cite{Heus06} 
\begin{equation}
\label{eq:45}
 \sum_{\vec{v},L-V=k}(-1)^{V} N(\vec{v}) = 
 \left(1-\frac{2}{\beta}\right)^{k} \ ,
\end{equation}
where $\beta=1$, if time reversal symmetry is present and $\beta=2$,
if time reversal symmetry is broken. 

Finally, using these results in the case of time-reversal invariance, we 
obtain for the full transmission matrix, including also the diagonal part, 
\begin{equation}
\label{eq:46}
 T_{a_2,a_1'}^{\mathrm{nd}}+\frac{2s+1}{N_1+N_2} \approx 
 \frac{(2s+1)^{2}}{(2s+1)(N_{1}+N_{2})-1} \ ,
\end{equation}
in the case of half-integer $s$, and 
\begin{equation}
\label{eq:47}
 T_{a_2,a_1'}^{\mathrm{nd}}+\frac{2s+1}{N_1+N_2} \approx
 \frac{(2s+1)^{2}}{(2s+1)(N_{1}+N_{2})+1} \ ,
\end{equation}
if $s$ is integer. For $s=1/2$ the result (\ref{eq:46}) is identical with 
the one obtained using Random Matrix Theory, in the circular symplectic 
ensemble~\cite{Bee97}. 

These findings can now be compared with the respective results when 
time-reversal is absent, thus revealing the behavior of the transmission
under a breaking of time-reversal by, e.g., turning on a magnetic field.
In that case $\beta =2$ so that the term (\ref{eq:45}) vanishes, implying
via (\ref{eq:43}) that only the diagonal contribution survives.
The difference $\Delta{\cal T}={\cal T}^{(\beta=1)}-{\cal T}^{(\beta=2)}$ 
of the transmission coefficients therefore is
\begin{equation}
\label{eq:48}
 \Delta{\cal T} \approx \frac{N_1 N_2 (2s+1)}{(N_{1}+N_{2})
 \left((2s+1)(N_{1}+N_{2})-1\right)} \ ,
\end{equation}
in the case of half-integer $s$, and
\begin{equation}
\label{eq:49}
 \Delta{\cal T} \approx \frac{-N_1 N_2 (2s+1)}{(N_{1}+N_{2})
 \left((2s+1)(N_{1}+N_{2})+1\right)}  \ ,
\end{equation}
if $s$ is integer. From these expressions one immediately concludes that
the transmission (i.e., conductivity) is enhanced at zero magnetic field
(when time reversal symmetry is restored), if the spin is half-integer;
thus weak anti-localization occurs. To the contrary, integer spin would
lead to weak localization. The latter had previously been obtained in 
semiclassical studies where the spin had been neglected~\cite{Sie02}. 
The only semiclassical derivation of weak anti-localization so 
far~\cite{Zai05}, however, was restricted to the one-loop contribution 
and employed asymptotics for large $N_1,N_2$.
\section{Shot noise}
The techniques developed above can be applied to a number of further
problems arising in the context of ballistic transport through chaotic
mesoscopic cavities. As a first example we consider shot noise. To this 
end one needs to compute the energy-averaged Fano-factor $F$, defined as
\begin{equation}
\label{eq:50}
 F := \frac{\left\langle {\mathrm{Tr}}(TT^{\dagger}-TT^{\dagger}TT^{\dagger})
      \right\rangle _{\Delta E}}{\left\langle {\mathrm{Tr}}(TT^{\dagger})
      \right\rangle _{\Delta E}} \ ,
\end{equation}
in terms of the transmission matrix $T=S^{21}$. The denominator has been 
dealt with above, and the spin-independent contribution to 
\begin{equation}
\label{eq:51}
 {\mathrm{Tr}}\left(TT^{\dagger}TT^{\dagger}\right) \ 
\end{equation}
was calculated semiclassically in Ref.~\onlinecite{Heu06}. We are hence 
left with the task of determining the spin contribution to (\ref{eq:51}).
Referring to the semiclassical representation (\ref{eq:12}) one immediately
realizes that a four-fold sum over classical trajectories emerges. In
addition to the case covered in Ref.~\onlinecite{Heu06} each term in this
sum acquires an additional factor of 
\begin{equation}
\label{eq:51a}
 {\mathrm{Tr}}\left(D_{s}^{\dagger}D_{u}D_{v}^{\dagger}D_{w}\right) \ ,
\end{equation}
in which the indices label the trajectories involved. The diagonal 
contribution to the four-fold sum occurs with $s=u$ and $v=w$, or with 
$s=w$ and $u=v$. In both cases unitarity implies
\begin{equation}
\label{eq:52}
 {\mathrm{Tr}}\left(D_{s}^{\dagger}D_{u}D_{v}^{\dagger}D_{w}\right)=
 2s+1 \ .
\end{equation}
Beyond this one has to consider the encounter of four trajectories. For 
the first time this has been done in quantum graphs~\cite{Sch03}, and has
later been extended in Ref.~\onlinecite{Heu06}. Following the method of
these papers, every trajectory consists of two parts, labeled by $1$ and $2$.
Approximately, one then has $s_{1}=w_{1}$, $u_{1}=v_{1}$, $s_{2}=u_{2}$
and $v_{2}=w_{2}$. Thus
\begin{eqnarray}
\label{eq:53}
 {\mathrm{Tr}}\left(D_{s}^{\dagger}D_{u}D_{v}^{\dagger}D_{w}\right) 
 &\approx& {\mathrm{Tr}}\left(D_{s_{1}}^{\dagger}D_{s_{2}}^{\dagger}D_{s_{2}}
           D_{v_{1}}D_{v_{1}}^{\dagger}D_{v_{2}}^{\dagger}D_{v_{2}}
           D_{s_{1}}\right)\nonumber \\
 & = & 2s+1 \ .
\end{eqnarray}
Following further the calculation of the Fano factor in 
Ref.~\onlinecite{Heu06}, we obtain 
\begin{equation}
\label{eq:54}
 F \approx \frac{N_{1}N_{2}}{(N_{1}+N_{2})^{2}} \ ,
\end{equation}
for $N_{1},N_{2}\gg 1$. This result coincides with the respective outcome
of a random matrix calculation in the symplectic ensemble~\cite{Bee97,Sav06}.
\section{Conductance fluctuations}
Universality of conductance fluctuations is often characterized in terms of 
the energy-averaged variance of ${\mathrm{Tr}}(TT^{\dagger})$. Instead of this 
quantity, the energy-averaged covariance of ${\mathrm{Tr}}(R^{n}R^{n\dagger})$, 
where $n=1,2$ labels the leads, can also be considered, see 
Ref.~\onlinecite{Bro05} for details. Our calculations are based on the first 
paper of Ref.~\onlinecite{Bro05}, whose method can still be applied when the 
Ehrenfest time is much smaller than the dwell time; this condition is fulfilled 
in the semiclassical limit considered here. 

The calculation of the variances again involves four-fold sums over 
trajectories, in which the spin contribution occurs in terms of the factors
\begin{equation}
\label{eq:55}
 {\mathrm{Tr}}\left(D_{s}D_{u}^{\dagger}\right)
 {\mathrm{Tr}}\left(D_{v}D_{w}^{\dagger}\right) \ .
\end{equation}
Switching off the spin-orbit interaction while preserving the presence of
spin~$s$, one obtains
\begin{equation}
\label{eq:56}
 {\mathrm{Tr}}\left(D_{s}D_{u}^{\dagger}\right)
 {\mathrm{Tr}}\left(D_{v}D_{w}^{\dagger}\right)=(2s+1)^{2} \  .
\end{equation}
In the presence of spin-orbit interaction one must examine the trajectories
involved more closely. Here we again consider the case $N_{1},\, N_{2}\gg1$.
The trajectories are divided into three parts labeled by $1$, $2$ and $3$,
and the relations $s_{1}=u_{1}$, $s_{2}=\overline{v}_{2}$, $s_{3}=u_{3}$,
$v_{1}=w_{1}$, $u_{2}=\overline{w}_{2}$, $v_{3}=w_{3}$ 
or $s_{1}=u_{1}$, $s_{2}=v_{2}$, $s_{3}=u_{3}$, $v_{1}=w_{1}$,
$u_{2}=w_{2}$, $v_{3}=w_{3}$ hold approximately. Here an over-bar indicates 
that these pieces are traversed in reverse direction. In the first case this
yields
\begin{equation}
\label{eq:57}
 {\mathrm{Tr}}\left(D_{s}D_{u}^{\dagger}\right) 
 {\mathrm{Tr}}\left(D_{v}D_{w}^{\dagger}\right) \approx 
 {\mathrm{Tr}}\left(D_{s_{2}}D_{u_{2}}^{\dagger}\right)
 {\mathrm{Tr}}\left(D_{u_{2}}D_{s_{2}}^{\dagger}\right) \ ,
\end{equation}
whereas in the second case
\begin{equation}
\label{eq:58}
 {\mathrm{Tr}}\left(D_{s}D_{u}^{\dagger}\right)
 {\mathrm{Tr}}\left(D_{v}D_{w}^{\dagger}\right) \approx 
 {\mathrm{Tr}}\left(D_{s_{2}}D_{u_{2}}^{\dagger}\right)^{2} \ .
\end{equation}
After an average over ${\mathrm{SU}}(2)$, very much alike in the main part 
of this work, we obtain for the first case~\cite{Bol03,Bol05}
\begin{equation}
\label{eq:59}
 \int_{SU(2)}\int_{SU(2)}dg_{a}dg_{b} {\mathrm{Tr}}\left(\pi_{s}\left(g_{a}
 g_{b}^{\dagger}\right)\right) {\mathrm{Tr}}\left(\pi_{s}\left(g_{b}
 g_{a}^{\dagger}\right)\right) = 1  \ , 
\end{equation}
and for the second case~\cite{Bol99}
\begin{equation}
\label{eq:60}
 \int_{SU(2)}\int_{SU(2)}dg_{a}dg_{b}
 \left[{\mathrm{Tr}}\left(\pi_{s}\left(g_{a}g_{b}^{\dagger}
 \right)\right)\right]^{2} = 1  \ .
\end{equation}
We follow Ref.~\onlinecite{Bro05} further and finally observe that, with
$N_{1},N_{2}\gg 1$, the energy-averaged variance of 
${\mathrm{Tr}}(TT^{\dagger})$ reads
\begin{equation}
\label{eq:61}
 \left\langle \text{var}\left({\mathrm{Tr}}\left(TT^{\dagger} \right)\right)
 \right\rangle_{\Delta E} \approx 2\,(2s+1)^{2}
 \frac{(N_{1}N_{2})^{2}}{(N_{1}+N_{2})^{4}} \ ,
\end{equation}
when the spin-orbit interaction is switched off, and 
\begin{equation}
\label{eq:62}
 \left\langle\text{var}\left( {\mathrm{Tr}}\left(TT^{\dagger} \right)\right)
 \right\rangle_{\Delta E} \approx 2\,\frac{(N_{1}N_{2})^{2}}
 {(N_{1}+N_{2})^{4}}
\end{equation}
in the presence of spin-orbit interaction. Again, this finding is in 
accordance with the respective result in the symplectic ensemble of 
RMT~\cite{Bee97,Sav06}. 
\section{Summary and Conclusions}
We considered the semiclassical description of ballistic transport through 
chaotic mesoscopic cavities in the presence of spin-orbit interactions. 
Our focus was the calculation of transmission coefficients. Here the principal
task was to verify the effect of weak anti-localization in the form
predicted by RMT.

Working within the framework of the Landauer formalism, our starting point 
was a semiclassical representation of Green functions for Hamiltonians
that contain a spin-orbit interaction. Transmission coefficients then require 
the evaluation of double sums over classical trajectories. The principal
difficulty presented by such expressions is to get hold of the interferences
thus occurring. This can be overcome successfully by exploiting the 
Sieber-Richter method, originally developed to perform analogous
calculations in the context of spectral fluctuations in classically
chaotic quantum systems.

We attacked the problem using the two established variants of the 
Sieber-Richter method: the configuration-space approach for the leading order, 
and the phase-space approach for the remaining contributions. In the first 
case a key input was a classical sum rule encoding an ergodic (and mixing)
behavior of the combined classical spin-orbit dynamics. Essential to the 
success of the phase-space approach was a calculation of the spin contribution 
to pairs of classical trajectories that are grouped together pairwise 
according to the structure of their almost self-encounters. This led to the 
central result given in Eq.~(\ref{eq:36}). The sign appearing points to the 
essential difference between the effects of half-integer spin as opposed to 
integer spin (including spin zero). This difference was then identified as
responsible for weak anti-localization or localization, respectively, to 
occur. We finally showed how our approach generalizes to semiclassical
descriptions of shot noise and of universal conductance fluctuations.  
\appendix
\section{Proof of the relation (\ref{eq:36})} 
We will show the validity of Eq.~(\ref{eq:36}) by induction with respect to 
the number $n$ of 2-encounters of two trajectories $\gamma\neq\gamma'$. 
The proof is based on the relations
\begin{equation}
\label{eq:37}
 \int_{SU(2)}dg {\mathrm{Tr}}\left(\pi_{s}(xgyg)\right)
 =\frac{(-1)^{2s}}{2s+1}{\mathrm{Tr}}\left(\pi_{s}(xy^{-1})\right)
\end{equation}
and
\begin{eqnarray}
\label{eq:38}
 \int_{SU(2)}\int_{SU(2)} dg \, dh &&
    {\mathrm{Tr}}\left(\pi_{s}(gwh^{-1}xg^{-1}yhz)\right) \nonumber \\
 &&= \frac{1}{(2s+1)^{2}} {\mathrm{Tr}}\left(\pi_{s}(yxwz)\right) ,
\end{eqnarray}
valid for all $w,\, x,\, y,\, z\in SU(2)$. For finite groups analogous 
identities have been shown in Ref.~\onlinecite{Bol05}; their proofs can be
directly carried over to the present case.

We now proceed in three steps:
\begin{enumerate}
\item 
First consider the case $n=0$, where $\gamma'=\gamma$. This also means
$\eta_{j}=1$ and $k_{j}=j$. Here we obtain
\begin{eqnarray}
\label{eq:39}
 M_{\gamma\gamma}
 & = & \int_{SU(2)}\ldots\int_{SU(2)}dg_{L}\ldots dg_{2} \nonumber \\
 &   & \times {\mathrm{Tr}}\left(\pi_{s}\left(g_{L}...g_{2}g_{2}^{\dagger}... 
       g_{L}^{\dagger}\right)\right) \nonumber \\
 & = & 2s+1 \ .
\end{eqnarray}
\item 
We assume the validity of (\ref{eq:36}) for two trajectories
$\gamma=\left(l_{1},a,b,l_{2},c,d,l_{3}\right)$ and 
$\gamma'=\left(l_{4},a,b,l_{5},c,d,l_{6}\right)$ as shown in 
Figure~\ref{cap:zwei}.
\begin{figure*}
\begin{center}
\includegraphics[width=15cm]{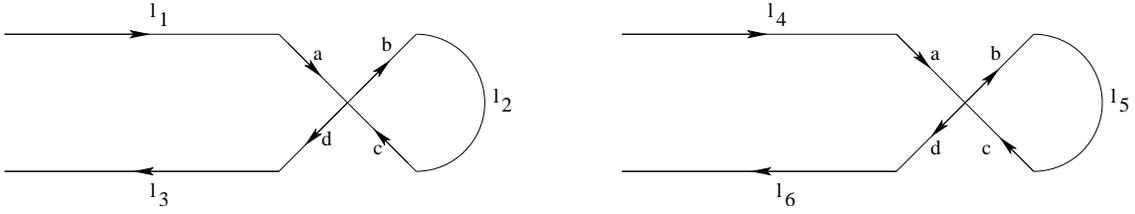}
\caption{Sketches of the trajectories $\gamma$ (left) and $\gamma'$ (right) 
that are considered under 2.}\label{cap:zwei}
\end{center}
\end{figure*}
Here $l_{j}$ stands for stretches of the trajectories $\gamma$ and $\gamma'$
containing an unspecified number of 2-encounters. By assumption, the actual 
number of 2-encounters, where $\gamma$ differs from $\gamma'$ is $n$.
We show now that the relation~(\ref{eq:36}) is still valid, when we replace
$\gamma'$ with the trajectory 
$\gamma''=\left(l_{4},a,\bar{c},\bar{l}_{5},\bar{b},d,l_{6}\right)$.
Thus $\gamma''$ differs from $\gamma$ in $n'=n+1$ 2-encounters. Then
\begin{widetext}
\begin{eqnarray}
\label{eq:40}
 M_{\gamma\gamma''}
 &=&\int_{SU(2)}\ldots\int_{SU(2)}dg_{a}dg_{b}dg_{c}dg_{d}\ldots
    {\mathrm{Tr}}\left(\pi_{s}\left(g_{l_{3}}g_{d}g_{c}g_{l_{2}}g_{b}g_{a}
    g_{l_{1}}g_{l_{4}}^{\dagger}g_{a}^{\dagger}g_{c}g_{l_{5}}g_{b}
    g_{d}^{\dagger}g_{l_{6}}^{\dagger}\right)\right)\nonumber \\
 &=&\int_{SU(2)}\ldots\int_{SU(2)}dg_{x}dg_{y}dg_{z}\ldots
    {\mathrm{Tr}}\left(\pi_{s}\left(g_{l_{3}}g_{x}g_{l_{2}}g_{y}g_{l_{1}}
    g_{l_{4}}^{\dagger}g_{y}^{\dagger}g_{z}g_{l_{5}}g_{z}g_{x}^{\dagger}
    g_{l_{6}}^{\dagger}\right)\right)\nonumber \\
 &=&\frac{(-1)^{2s}}{2s+1}\int_{SU(2)}\ldots\int_{SU(2)}dg_{x}dg_{y}\ldots
    {\mathrm{Tr}}\left(\pi_{s}\left(g_{l_{3}}g_{x}g_{l_{2}}g_{y}g_{l_{1}}
    g_{l_{4}}^{\dagger}g_{y}^{\dagger}g_{l_{5}}^{\dagger}g_{x}^{\dagger}
    g_{l_{6}}^{\dagger}\right)\right)\nonumber \\
 &=&\frac{(-1)^{2s}}{2s+1}\int_{SU(2)}\ldots\int_{SU(2)}dg_{a}dg_{b}dg_{c}
    dg_{d}\ldots {\mathrm{Tr}}\left(\pi_{s}\left(g_{l_{3}}g_{d}g_{c}g_{l_{2}}
    g_{b}g_{a}g_{l_{1}}g_{l_{4}}^{\dagger}g_{a}^{\dagger}g_{b}^{\dagger}
    g_{l_{5}}^{\dagger}g_{c}^{\dagger}g_{d}^{\dagger}
    g_{l_{6}}^{\dagger}\right)\right)\nonumber \\
 &=&\frac{(-1)^{2s}}{2s+1} \, M_{\gamma\gamma'} \ .
\end{eqnarray}
\end{widetext}
In the second step we substituted $g_{d}g_{c}=g_{x}$, $g_{b}g_{c}=g_{z}$
and $g_{b}g_{a}=g_{y}$, and in the third one we used Eq.~(\ref{eq:37}).
In the fourth step we undid the substitution. This calculation 
proves that changing the number of 2-encounters, in which $\gamma$ and 
$\gamma'$ differ, by one indeed contributes a factor of $(-1)^{2s}/(2s+1)$.
\item 
We assume the validity of the relation (\ref{eq:36}) for the two
trajectories $\gamma=\left(l_{1},a_{1},b_{1},l_{2},a_{2},b_{2},l_{3},c_{1},
d_{1},l_{4},c_{2},d_{2},l_{5}\right)$ and $\gamma'=\left(l_{6},a_{1},b_{1},
l_{7},a_{2},b_{2},l_{8},c_{1},d_{1},l_{9},c_{2},d_{2},l_{10}\right)$ as
shown in Figure~\ref{cap:drei}.
\begin{figure*}
\begin{center}\includegraphics[%
  width=14cm]{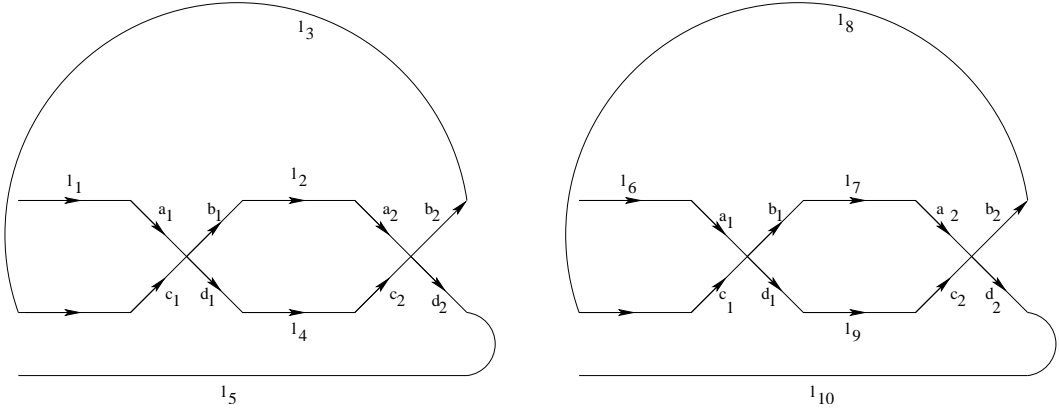}
\caption{Sketches of the trajectories $\gamma$ (left) and $\gamma'$ (right)
that are considered under 3.}\label{cap:drei}
\end{center}
\end{figure*}
Again we assume that the number of 2-encounters, where $\gamma$ differs from 
$\gamma'$, is $n$. We then show that the relation~(\ref{eq:36}) is unchanged
under a replacement of $\gamma'$ with the trajectory 
$\gamma''=\left(l_{6},a_{1},d_{1},l_{9},c_{2},b_{2},l_{8},c_{1},b_{1},
l_{7},a_{2},d_{2},l_{10}\right)$. Notice that $\gamma''$ cannot be constructed
by applying the procedure of 2. twice: here the stretches $l_{6}$, $l_{7}$ 
and $l_{9}$ of $\gamma'$ are traversed in parallel direction, whereas in 2.
the stretches $l_{4}$ and $l_{6}$ of $\gamma'$ are traversed in anti-parallel 
direction. A calculation similar to (\ref{eq:40}), with the substitutions 
$g_{d_j}g_{c_j}=g_{x_j}$, $g_{b_j}g_{d_j}^{\dagger}=g_{z_j}$, 
$g_{b_j}g_{a_j}=g_{y_j}$ $\left( j\in\left\lbrace  1,2\right\rbrace\right)$, 
then yields
\begin{widetext}
\begin{eqnarray}
\label{eq:41}
 M_{\gamma\gamma''}
 &=&\int_{SU(2)}\ldots\int_{SU(2)}dg_{a_{1}}\ldots{\mathrm{Tr}}\left(\pi_{s}
    \left(g_{l_{5}}g_{d_{2}}g_{c_{2}}g_{l_{4}}g_{d_{1}}g_{c_{1}}g_{l_{3}}
    g_{b_{2}}g_{a_{2}}g_{l_{2}}g_{b_{1}}g_{a_{1}}g_{l_{1}}g_{l_{6}}^{\dagger}
    g_{a_{1}}^{\dagger}g_{d_{1}}^{\dagger}g_{l_{9}}^{\dagger}
    g_{c_{2}}^{\dagger}g_{b_{2}}^{\dagger}g_{l_{8}}^{\dagger}
    g_{c_{1}}^{\dagger}g_{b_{1}}^{\dagger}g_{l_{7}}^{\dagger}
    g_{a_{2}}^{\dagger}g_{d_{2}}^{\dagger}g_{l_{10}}^{\dagger}\right)\right)
    \nonumber \\
 &=&\int_{SU(2)}\ldots\int_{SU(2)}dg_{x_{1}}\ldots{\mathrm{Tr}}\left(\pi_{s}
    \left(g_{l_{5}}g_{x_{2}}g_{l_{4}}g_{x_{1}}g_{l_{3}}g_{y_{2}}g_{l_{2}}
    g_{y_{1}}g_{l_{1}}g_{l_{6}}^{\dagger}g_{y_{1}}^{\dagger}g_{z_{1}}
    g_{l_{9}}^{\dagger}g_{x_{2}}^{\dagger}g_{z_{2}}^{\dagger}
    g_{l_{8}}^{\dagger}g_{x_{1}}^{\dagger}g_{z_{1}}^{\dagger}
    g_{l_{7}}^{\dagger}g_{y_{2}}^{\dagger}g_{z_{2}}g_{l_{10}}^{\dagger}\right)
    \right)\nonumber \\
 &=&\frac{1}{(2s+1)^{2}}\int_{SU(2)}\ldots\int_{SU(2)}dg_{x_{1}}
    dg_{y_{1}}dg_{x_{2}}dg_{y_{2}}\ldots{\mathrm{Tr}}\left(\pi_{s}\left(
    g_{l_{7}}^{\dagger}g_{y_{2}}^{\dagger}g_{l_{8}}^{\dagger}
    g_{x_{1}}^{\dagger}g_{l_{9}}^{\dagger}g_{x_{2}}^{\dagger}
    g_{l_{10}}^{\dagger}g_{l_{5}}g_{x_{2}}g_{l_{4}}g_{x_{1}}g_{l_{3}}g_{y_{2}}
    g_{l_{2}}g_{y_{1}}g_{l_{1}}g_{l_{6}}^{\dagger}g_{y_{1}}^{\dagger}\right)
    \right)\nonumber \\
 &=&\frac{1}{(2s+1)^{2}} \, M_{\gamma\gamma'} \ .
\end{eqnarray}
\end{widetext}
\end{enumerate}
After these steps (\ref{eq:36}) follows by induction because every trajectory 
$\gamma'$ can be constructed successively out of $\gamma$ by using the 
procedures of 2. and 3. Every $l$-encounter that does not decompose into 
several encounters of a lower number of trajectories (see Figure~4 in 
Ref.~\onlinecite{Heu05} for an example) can be constructed from 2-encounters
in $l-1$ steps. Every such step then brings out a factor of 
$(-1)^{2s}/(2s+1)$ in $M_{\gamma,\gamma'}$, when this is constructed from 
$M_{\gamma,\gamma}=2s+1$. Thus, $V$ encounters with altogether $L$ stretches
contribute a factor $\left((-1)^{2s}/(2s+1)\right)^{L-V}$, which completes 
the proof of (\ref{eq:36}).
%
%
%
%
%
%


\begin{thebibliography}{10}
%
\bibitem{Bee97}
C.W.J. Beenakker, 
\emph{Random-matrix theory of quantum transport},
Rev. Mod. Phys. \textbf{69} 731 (1997)
%
\bibitem{Lan57} 
R. Landauer, 
\emph{Spatial variation of currents and fields due to localized scatterers 
in metallic conduction}, 
IBM J. Res. Develop. \textbf{1} 223 (1957); 
IBM J. Res. Develop. \textbf{32} 306 (1988) 
%
\bibitem{Bar93}
H. U. Baranger, R. A. Jalabert, A. D. Stone, 
\emph{Quantum-chaotic scattering effects in semiconductor microstructures}, 
Chaos \textbf{3} 665 (1993); 
\emph{Weak Localization and Integrability in Ballistic Cavities}, 
Phys. Rev. Lett. \textbf{70} 3876 (1993)
%
\bibitem{Ric01}
M. Sieber, K. Richter, 
\emph{Correlations between Periodic Orbits and their R\^{o}le in Spectral 
Statistics}, 
Physica Scripta \textbf{T90} 128 (2001) 
%
\bibitem{Mue04} 
S. M\"uller, S. Heusler, P. Braun, F. Haake, A. Altland, 
\emph{Semiclassical Foundation of Universality in Quantum Chaos}, 
Phys. Rev. Lett. \textbf{93} 014103 (2004)  
%
\bibitem{Heu05} 
S. Heusler, S. M\"uller, P. Braun, F. Haake, A. Altland,
\emph{Periodic-orbit theory of universality in quantum chaos}, 
Phys. Rev. E \textbf{72} 046207 (2005) 
%
\bibitem{He07} 
S. Heusler, S. M\"uller, A. Altland, P. Braun, F. Haake, 
\emph{Periodic-Orbit Theory of Level Correlations}, 
Phys. Rev. Lett. \textbf{98} 044103 (2007)
%
\bibitem{Sie02}
K. Richter, M. Sieber, 
\emph{Semiclassical Theory of Chaotic Quantum Transport}, 
Phys. Rev. Lett. \textbf{89} 206801 (2002) 
%
\bibitem{Heus06}
S. Heusler, S. M\"uller, P. Braun, F. Haake, 
\emph{Semiclassical Theory of Chaotic Conductors}, 
Phys. Rev. Lett. \textbf{96} 066804 (2006) 
%
\bibitem{Heu07}
S. M\"uller, S. Heusler, P. Braun, F. Haake,
\emph{Semiclassical approach to chaotic quantum transport}, 
New J. Phys. \textbf{9} 12 (2007) 
%
\bibitem{Bro05}
P.W. Brouwer, S. Rahav, 
\emph{A semiclassical theory of the Ehrenfest-time dependence of quantum 
transport in ballistic quantum dots}, 
preprint, arXiv:cond-mat/0512095v2 (2005); 
Phys. Rev. B \textbf{74} 075322 (2006)
%
\bibitem{Ber07}
G. Berkolaiko, J.M. Harrison, M. Novaes,
\emph{Full counting statistics of chaotic cavities from classical action
correlations},
preprint, arXiv:cond-mat/0703803v2 (2007)
%
\bibitem{Sch03}
H. Schanz, M. Puhlmann, T. Geisel, 
\emph{Shot Noise in Chaotic Cavities from Action Correlations}, 
Phys. Rev. Lett. \textbf{91} 134101 (2003)
%
\bibitem{Heu06}
S. Heusler, S. M\"uller, P. Braun, F. Haake, 
\emph{Semiclassical Prediction for Shot Noise in Chaotic Cavities}, 
J. Phys. A: Math. Gen. \textbf{39} L159 (2006)
%
\bibitem{Fab04}
I. \v{Z}uti\'{c}, J. Fabian, S. Das Sarma, 
\emph{Spintronics: Fundamentals and applications}, 
Rev. Mod. Phys. \textbf{76} 323 (2004) 
%
\bibitem{Sav06}
D. V. Savin, H.-J. Sommers, 
\emph{Shot noise in chaotic cavities with an arbitrary number of open 
channels}, 
Phys. Rev. B \textbf{73} 081307(R) (2006) 
%
\bibitem{Zai05}
O. Zaitsev, D. Frustaglia, K. Richter, 
\emph{Role of orbital dynamics in spin relaxation and weak antilocalization
in quantum dots},
Phys. Rev. Lett. \textbf{94} 026809 (2005);
\emph{Semiclassical theory of weak antilocalization and spin relaxation in 
ballistic quantum dots}, 
Phys. Rev. B \textbf{72} 155325 (2005)
%
\bibitem{Bol99}
J. Bolte, S. Keppeler, 
\emph{A semiclassical approach to the Dirac equation}, 
Ann. Phys. (NY) \textbf{274} 125 (1999) 
%
\bibitem{Bol99a}
J. Bolte, S. Keppeler, 
\emph{Semiclassical form factor for chaotic systems with spin 1/2}, 
J. Phys. A: Math. Gen. \textbf{32} 8863 (1999)
%
\bibitem{Bol03}
J. Bolte, J. Harrison, 
\emph{The spin contribution to the form factor of quantum graphs}, 
J. Phys. A: Math. Gen. \textbf{36} L433 (2003)
%
\bibitem{Bol05}
J. Bolte, J. Harrison, 
\emph{The spectral form factor for quantum graphs with spin-orbit coupling}, 
in: G. Berkolaiko, R. Carlson, S.A. Fulling, and P. Kuchment (eds.): 
\emph{Quantum Graphs and Their Applications}, 
Contemporary Mathematics, Volume {\bf 415}, pp. 51 (AMS 2006)
%
\bibitem{Fis81} 
D.S. Fisher, P.A. Lee, 
\emph{Relation between conductivity and transmission matrix}, 
Phys. Rev. B \textbf{23} 6851 (1981) 
%
\bibitem{Bar89}
H.U. Baranger, A.D. Stone, 
\emph{Electrical} \emph{linear-response theory in an arbitrary magnetic field: 
A new Fermi-surface formation}, 
Phys. Rev. B, \textbf{40} 8169 (1989) 
%
\bibitem{Noe93}
J.U. N\"ockel, A.D. Stone, H.U. Baranger, 
\emph{Adiabatic turn-on and the asymptotic limit in linear-response theory 
for open systems},
Phys. Rev. B \textbf{48} 17569 (1993) 


\end{thebibliography}
\end{document}